\documentclass[twocolumn,showpacs,preprintnumbers,amsmath,amssymb,nofootinbib]{revtex4-1}

\usepackage[dvipdfmx]{graphicx}
\usepackage{amsfonts}
\usepackage[mathscr]{eucal}
\usepackage{dcolumn}
\usepackage{bm}
\usepackage[colorlinks=true,linkcolor=blue,citecolor=blue]{hyperref}
\usepackage{hyperref}
\usepackage{color}

\begin{document}


\newcommand{\vev}[1]{ \left\langle {#1} \right\rangle }
\newcommand{\bra}[1]{ \langle {#1} | }
\newcommand{\ket}[1]{ | {#1} \rangle }
\newcommand{\eV}{ \ {\rm eV} }
\newcommand{\KeV}{ \ {\rm keV} }
\newcommand{\MeV}{\  {\rm MeV} }
\newcommand{\GeV}{\  {\rm GeV} }
\newcommand{\TeV}{\  {\rm TeV} }
\newcommand{\1}{\mbox{1}\hspace{-0.25em}\mbox{l}}
\newcommand{\Red}[1]{{\color{red} {#1}}}

\newcommand{\lmk}{\left(}  
\newcommand{\rmk}{\right)}
\newcommand{\lkk}{\left[}  
\newcommand{\rkk}{\right]}
\newcommand{\lhk}{\left \{ }  
\newcommand{\rhk}{\right \} }
\newcommand{\del}{\partial}  
\newcommand{\la}{\left\langle} 
\newcommand{\ra}{\right\rangle}
\newcommand{\half}{\frac{1}{2}}

\newcommand{\bea}{\begin{array}}
\newcommand{\eea}{\end{array}}
\newcommand{\beq}{\begin{eqnarray}}
\newcommand{\eeq}{\end{eqnarray}}

\newcommand{\dd}{\mathrm{d}}
\newcommand{\Mpl}{M_{\rm Pl}}
\newcommand{\mg}{m_{3/2}}
\newcommand{\abs}[1]{\left\vert {#1} \right\vert}
\newcommand{\mphi}{m_{\phi}}
\newcommand{\Hz}{\ {\rm Hz}}
\newcommand{\for}{\quad \text{for }}
\newcommand{\Min}{\text{Min}}
\newcommand{\Max}{\text{Max}}
\newcommand{\Kahler}{K\"{a}hler }
\newcommand{\cphi}{\varphi}
\newcommand{\Tr}{\text{Tr}}
\newcommand{\diag}{{\rm diag}}

\newcommand{\SUf}{SU(3)_{\rm f}}
\newcommand{\Upq}{U(1)_{\rm PQ}}
\newcommand{\Zpq}{Z^{\rm PQ}_3}
\newcommand{\Cpq}{C_{\rm PQ}}
\newcommand{\ubar}{u^c}
\newcommand{\dbar}{d^c}
\newcommand{\ebar}{e^c}
\newcommand{\nubar}{\nu^c}
\newcommand{\Ndw}{N_{\rm DW}}
\newcommand{\Fpq}{F_{\rm PQ}}
\newcommand{\fpq}{v_{\rm PQ}}
\newcommand{\Br}{{\rm Br}}
\newcommand{\Lag}{\mathcal{L}}
\newcommand{\Lqcd}{\Lambda_{\rm QCD}}
\newcommand{\const}{\text{const}}

\newcommand{\ji}{j_{\rm inf}} 
\newcommand{\jb}{j_{B-L}} 
\newcommand{\M}{M} 
\newcommand{\im}{{\rm Im} }
\newcommand{\re}{{\rm Re} }
\newcommand{\cm}{\ {\rm cm} }

\def\lrf#1#2{ \left(\frac{#1}{#2}\right)}
\def\lrfp#1#2#3{ \left(\frac{#1}{#2} \right)^{#3}}
\def\lrp#1#2{\left( #1 \right)^{#2}}
\def\REF#1{Ref.~\cite{#1}}
\def\SEC#1{Sec.~\ref{#1}}
\def\FIG#1{Fig.~\ref{#1}}
\def\EQ#1{Eq.~(\ref{#1})}
\def\EQS#1{Eqs.~(\ref{#1})}
\def\blue#1{\textcolor{blue}{#1}}
\def\red#1{\textcolor{blue}{#1}}

\newcommand{\fa}{f_{a}}
\newcommand{\Uh}{U(1)$_{\rm H}$}
\newcommand{\osc}{_{\rm osc}}

\newcommand{\mav}{\left. m_a^2 \right\vert_{T=0}}
\newcommand{\mat}{m_{a, {\rm QCD}}^2 (T)}
\newcommand{\mam}{m_{a, {\rm M}}^2 }
\def\eq#1{Eq.~(\ref{#1})}

\newcommand{\LQCD}{\Lambda_{\rm QCD}}

\newcommand{\UH}{U(1)$_H$ }

\newcommand{\EV}{ \ {\rm eV} }
\newcommand{\KEV}{ \ {\rm keV} }
\newcommand{\MEV}{\  {\rm MeV} }
\newcommand{\GEV}{\  {\rm GeV} }
\newcommand{\TEV}{\  {\rm TeV} }

\def\order#1{\mathcal{O}(#1)}


\preprint{
TU-1099; \ 
MIT-CTP/5186 
}

\title{
Strongly-interacting massive particle and dark photon \\
in the era of intensity frontier 
}

\author{
Ayuki Kamada,$^{1}$
}

\author{
Masaki Yamada,$^{2, 3, 4}$
}

\author{
Tsutomu T. Yanagida$^{5, 6}$
}

\affiliation{
$^{1}$ Center for Theoretical Physics of the Universe, 
Institute for Basic Science (IBS), 55 Expo-ro, Yuseong-gu, Daejeon 34126, Korea
}
\affiliation{
$^{2}$ Frontier Research Institute for Interdisciplinary Sciences, Tohoku University, Sendai, Miyagi 980-8578, Japan
}
\affiliation{
$^{3}$ Department of Physics, Tohoku University, Sendai, Miyagi 980-8578, Japan
}
\affiliation{
$^{4}$ Center for Theoretical Physics, Laboratory for Nuclear Science and Department of Physics, 
\\
Massachusetts Institute of Technology, Cambridge, MA 02139, USA 
}

\affiliation{
$^{5}$ T. D.  Lee Institute and School of Physics and Astronomy, Shanghai Jiao Tong University, 800 Dongchuan Rd, Shanghai 200240, China
}
\affiliation{$^{6}$ Kavli IPMU (WPI), UTIAS, 
The University of Tokyo, 5-1-5 Kashiwanoha, Kashiwa, Chiba 277-8583, Japan
}

\date{\today}

\begin{abstract} 
A strongly interacting massive particle (SIMP) is an interesting candidate for dark matter (DM) because its self-interaction cross section can be naturally strong enough to address the astrophysical problem of small-scale structure formation. A simple model was proposed by assuming a monopole condensation, where composite SIMP comes from a ``strongly interacting" U(1)$_{\rm d}$ gauge theory. In the original model, the DM relic abundance is determined by the $3\to2$ annihilation process via the Wess-Zumino-Witten term. In this paper, we discuss that the DM relic abundance is naturally determined also by a semi-annihilation process via a kinetic mixing between the hypercharge gauge boson and the dark U(1)$_{\rm d}$ gauge boson (dark photon). The dark photon can be discovered by LDMX-style missing momentum experiments in the near future. 
\end{abstract}

\maketitle

\section{Introduction
\label{sec:introduction}}
The intensity frontier is one of the broad approaches 
to new physics in collider experiments 
and recently became more important 
as the Large Hadron Collider has not yet found a clear signal for new physics. 
We should also note the null results in direct-detection experiments of dark matter (DM), 
which may indicate that the mass of DM is not of order the electroweak or TeV scale. 
We therefore focus on the case in which the DM mass is in a sub-GeV region, 
which 
can be tested via rare events rather than by a direct production from high-energy particles. 
Among proposed high-intensity accelerators, 
the Light Dark Matter eXperiment (LDMX)~\cite{Akesson:2018vlm} is designed to measure missing momentum 
in high-rate electron fixed-target reactions and can be a powerful discovery tool for such a light DM particle.

From the perspective of cosmology, 
the strongly-interacting massive particle (SIMP) 
proposed in Refs.~\cite{Hochberg:2014dra, Hochberg:2014kqa} naturally fits sub-GeV DM. 
They pointed out that 
the relic abundance of sub-GeV DM is consistent with the observed value 
if the $3 \to 2$ annihilation process dominates at the time of the freeze-out of DM 
and its cross section is determined by the mass scale of DM with an ${\cal O}(1)$ coupling. 
SIMPs can be naturally realized by composite particles like pions. 
The $3 \to 2$ annihilation process is actually realized by the Wess-Zumino-Witten term 
in the low-energy dark sector. 
Interestingly, 
the model predicts a self-interaction cross section of DM 
which is potentially favored by the observations of small-scale structure in cosmology~\cite{Spergel:1999mh, deBlok:2009sp, BoylanKolchin:2011de, Zavala:2012us, Kamada:2016euw} (see Ref.~\cite{Tulin:2017ara} for a review). 
This is dubbed as the SIMP miracle. 
However, 
there is a difficulty in maintaining thermal equilibrium between the dark and visible sectors 
during the freeze-out of the $3 \to 2$ annihilation process, 
which is required for the SIMP miracle to work. 
This can be realized in rather complicated models like the ones proposed in Refs.~\cite{Lee:2015gsa, Hochberg:2015vrg, Bernal:2015ova} (see Refs.~\cite{Tsumura:2017knk, 
Kamada:2017tsq, Ho:2017fte, Matsuzaki:2017bpp, Choi:2017zww, Berlin:2018tvf, Choi:2018iit, Heikinheimo:2018esa, Hochberg:2018rjs} for recent works).

In Ref.~\cite{Kamada:2016ois}, we have proposed a simple model of the SIMP, 
where the composite DM ``pions" consist of dark-sector ``electrons" and ``positrons" connected by a U(1)$_{\rm d}$ gauge interaction 
rather than a strong non-Abelian gauge interaction. 
We introduce a fundamental ``monopole" for U(1)$_{\rm d}$ at a high-energy scale and 
assume a ``monopole" condensation at the sub-GeV scale. 
One cannot write down the Lagrangian of this kind of theory including both a ``monopole" and an ``electron". However, this does not mean that the theory does not exist. In fact, theories with ``monopoles" and ``electrons" have been extensively studied in ${\cal N} = 2$~\cite{Seiberg:1994rs, Seiberg:1994aj, Argyres:1995jj, Argyres:1995xn} and ${\cal N} = 1$ supersymmetry~\cite{Bolognesi:2015wta, Giacomelli:2014rna, Xie:2016hny, Buican:2016hnq} 
without specifying the Lagrangian. 
In this paper, we revisit our SIMP model and propose a scenario in which 
the DM relic abundance is determined by a $2 \to 2$ semi-annihilation process 
via the kinetic mixing between the U(1)$_{\rm d}$ gauge boson and U(1)$_Y$ gauge boson 
rather than the $3 \to 2$ annihilation process. 
The model is quite economical~\cite{Yamada:2015waa}; 
we do not need to introduce any other particles 
but just introduce dark-sector ``electrons", a ``monopole", and the U(1)$_{\rm d}$ gauge boson (dark photon), 
the latter of which plays the roles of confinement and mediator to the visible sector. 
Although the SIMP miracle does not work in this scenario, 
the model is simple 
and all small dimensionless parameters are expected to be naturally small due to non-trivial anomalous dimensions.

The detectability and testability of our model is quite different from other DM models. 
Since there is no ``pion"-``pion"-photon interaction 
and the semi-annihilation process is $p$-wave suppressed, 
it is very difficult to directly or indirectly detect the DM ``pions". 
However, the kinetic mixing allows us to discover the dark photon by LDMX-like experiments. 
Our model is unique in the sense that it can be tested mainly by experiments designed to measuring missing momentum in high-rate electron fixed-target reactions.

The organization of this paper is as follows. 
In the next section, we specify particle contents of our model at high- and low-energy scales. 
We assume that the U(1)$_{\rm d}$ gauge symmetry is spontaneously broken by a ``monopole'' condensation at the energy scale of $0.1$-$1 \GeV$, below which 
there are ``pions". 
We calculate its self-interaction cross section 
and show that it is within the value potentially favored by the observations of small-scale structure. 
In Sec.~\ref{sec:3}, we explain how the ``pion" relic abundance is determined by the freeze-out process, 
taking into account a kinetic mixing between U(1)$_{\rm d}$ and U(1)$_Y$ gauge bosons. 
The relevant process is a semi-annihilation, 
which shows the freeze-out qualitatively different but is quantitatively similar to the standard freeze-out via annihilation. 
We take all ${\cal O}(1)$ parameters to be within $(0.1,1)$ 
for a conservative calculation 
and present a consistent parameter space for the kinetic mixing parameter 
and the mass of the U(1)$_{\rm d}$ gauge boson. 
Then we discuss the condition that 
the $3\to2$ annihilation process is negligible in our calculation. 
Finally we comment on the mixing between the SM Higgs and the ``monopole". 
Sec.~\ref{conclusions} is devoted to conclusions.

\section{Hidden ``pions" from a ``monopole" condensation
\label{sec:model}}
We introduce a scalar ``monopole" $\phi$ and $N_F$ pairs of dark-sector ``electrons" $\psi_i$ and ``positrons" $\bar{\psi}_i$ with U(1)$_{\rm d}$ gauge field~\cite{Kamada:2016ois}. 
To ensure the stability of ``pions" in the low-energy dark sector, 
we assume SU($N_F$) flavor symmetry under which the ``electrons" and ``positrons" 
transform in the fundamental and anti-fundamental representations, respectively. 
The charge assignment for $\psi_i$ and $\bar{\psi}_i$ is summarized in Table~\ref{table1}. 
We call the U(1)$_{\rm d}$ gauge boson as a dark photon. 

\begin{table}\begin{center}
\begin{tabular}{|p{1.0cm}|p{1.5cm}|p{1.5cm}|p{1.5cm}|p{1.5cm}|}
  \hline
  \rule[-5pt]{0pt}{15pt}
& \hfil SU($N_F$) \hfil 
    & \hfil U(1)$_{\rm d}$ \hfil & \hfil U(1)$_Y$ \hfil  \\
  \hline
  \rule[-5pt]{0pt}{15pt}
  \hfil $\psi_i$ \hfil 
& \hfil $\Box$ \hfil 
  & \hfil $1$ \hfil & \hfil 0 \hfil  \\
  \hline
  \rule[-5pt]{0pt}{15pt}
  \hfil $\bar{\psi}_i$ \hfil 
& \hfil $\bar{\Box}$ \hfil 
  & \hfil $-1$ \hfil & \hfil 0 \hfil  \\
\hline
\end{tabular}\end{center}
\caption{Charge assignment for matter fields in the dark sector.
\label{table1}}
\end{table}

We consider the case where 
the U(1)$_{\rm d}$ gauge symmetry is spontaneously broken by the ``monopole" condensation 
in the low-energy dark sector, just like the Higgs mechanism~\cite{Nambu:1974zg}. 
Each pair of ``electrons" and ``positrons" is then confined and connected by a string formed by the ``monopole" condensation~\cite{Nambu:1974zg}
and composes mesons
while there is no baryon state in the low-energy dark sector~\cite{Yamada:2016jgg}. 
The string tension is determined by the energy scale of the ``monopole" condensation, $\Lambda$, and sets the dynamical scale of the system. 
We assume the condensation of ``electrons" and ``positrons" that dynamically breaks the chiral symmetry and the ``pions" are the lightest composite states in the low-energy dark sector. 
We also assume that the chiral symmetry for the ``electrons" and ``positrons" is only an approximate symmetry so that the mass of the ``pions" is as large as (but smaller than) the condensation scale $\Lambda$~\cite{Hochberg:2014kqa, Hansen:2015yaa}.

After the ``monopole" condensation, 
there are $N_\pi = N_F^2 - 1$ ``pions", the radial component of ``monopole", 
and a massive U(1)$_{\rm d}$ gauge boson in the effective field theory. 
The ``monopole" and the gauge boson are assumed to be heavier than the ``pions", 
which we identify as DM. 

There is only one energy scale in the dark sector $\Lambda$, 
which is of order the masses of ``pions", ``monopole", and dark photon denoted by $m_\pi$, $m_\phi$, and $m_V$, respectively. 
We introduce ${\cal O}(1)$ constants $c_i$ 
that represents our ignorance of an ${\cal O}(1)$ uncertainty in the low-energy effective field theory~\cite{Hansen:2015yaa}. 
For example, we define 
$m_\pi = c_\Lambda \Lambda = c_{m_\phi} m_\phi = c_{m_V} m_V$. 
We also introduce other ${\cal O}(1)$ parameters associated with interactions in the dark sector specified below. 
To calculate the conservative bounds, we take $c_i \in (0.1, 1)$ throughout this paper.%
\footnote{
In Ref.~\cite{Kamada:2016ois}, 
we assumed $c_i = 1$ for simplicity. 
However, these uncertainties are important to discuss the detectability of our model in collider experiments, like LDMX. 
}

\subsection{Self-interactions}
\label{sec:2-2}

The ``pions" have self-interactions whose cross sections are determined by the size of ``pions", which is of order $\Lambda^{-1}$. 
Representing an ${\cal O}(1)$ factor by $c_1$, 
we write the cross section as 
\beq
 \frac{\sigma_{\rm ela}}{m_\pi} &=& \frac{(4 \pi)^4 c_1^2 m_\pi}{4 \pi \Lambda^4} 
 \nonumber \\
 &\simeq& 2.7 \cm^2 / {\rm g} \  \lmk \frac{c_1 c_\Lambda^{2}}{(4 \pi)^{-1}} \rmk^2 
 \lmk \frac{m_\pi}{100 \MeV} \rmk^{-3}. 
 \label{sigma_ela}
\eeq
from the dimensional analysis.%
\footnote{
\label{footnote1}
We assume $c_1 c_{\Lambda}^2 \lesssim (4 \pi)^{-1}$ throughout this paper so that 
the scattering cross section is less than the geometrical cross section, $4 \pi / m_\pi^2$, that is below the Unitarity bound for $v < c$~\cite{Griest:1989wd}. 
}
This is of order the upper bound 
on the self-interaction cross section of DM from 
the observations of cluster collisions, including the bullet cluster, 
and 
ellipticity on Milky way and cluster scales~\cite{Clowe:2003tk, Markevitch:2003at, Randall:2007ph, Rocha:2012jg, Peter:2012jh}. 
These 
constraints and discussions 
have ${\cal O}(1)$ uncertainties 
due to, say, the difficulties of numerical simulations, 
and hence we consider that 
they are marginally consistent with $\sigma_{\rm ela} / m_\pi = 0.1-1 \cm^2 / {\rm g}$. 
The recent observations of small-scale structure potentially 
favors the self-interacting DM with a cross section of the same order~\cite{Spergel:1999mh, deBlok:2009sp, BoylanKolchin:2011de, Zavala:2012us, Kahlhoefer:2015vua, Kamada:2016euw}. 
We note that $m_\pi$ can be as small as about $10 \MeV$ if $c_\Lambda = c_1 = 0.1$.

\section{Relic abundance of ``pions"}
\label{sec:3}

\subsection{Kinetic mixings and $2 \to 2$ semi-annihilation process}
There must be a nonzero kinetic mixing $\epsilon$ between 
the U(1)$_{\rm d}$ gauge boson and the U(1)$_Y$ gauge boson because it is allowed by any symmetry~\cite{Yamada:2016jgg}. 
There are two types of kinetic mixing terms in theories consisting simultaneously of both a ``monopole" and an ``electron": 
$\epsilon' B_{\mu \nu} F^{\mu \nu}$ and $\epsilon B_{\mu \nu} \tilde{F}^{\mu \nu}$, 
where 
$B_{\mu \nu}$ and $F_{\mu \nu}$ are the field strengths of U(1)$_Y$ and U(1)$_{\rm d}$ gauge bosons, respectively, 
and $\tilde{F}^{\mu \nu} \equiv (1/2) \varepsilon^{\mu \nu \rho \sigma} F_{\rho \sigma}$. 
If the CP symmetry is conserved, either of these mixing terms is allowed.%
\footnote{
One may think that $\epsilon B_{\mu \nu} \tilde{F}^{\mu \nu}$ itself violates the CP symmetry. 
In general, either of $F_{\mu \nu}$ and $\tilde{F}_{\mu \nu}$ can be chosen to be a tensor 
and the other one is a pseudo-tensor. 
If we choose the definition in which $B_{\mu \nu}$ and $\tilde{F}_{\mu \nu}$ are tensors 
and $\tilde{B}_{\mu \nu}$ and $F_{\mu \nu}$ are pseudo-tensors, 
the kinetic mixing term $\epsilon B_{\mu \nu} \tilde{F}^{\mu \nu}$ conserves the CP symmetry. 
In this case, dark ``pions" transform as $\pi \to - \pi$ (rather than $\pi \to - \pi^{\rm T}$) under the CP, 
so that $({\rm Tr} \left[ \pi \del_\mu \pi \del_\nu \pi \right] - (\mu \leftrightarrow \nu))$ is also a tensor 
and can be mixed with $\tilde{F}_{\mu \nu}$. 
}
However, one may expect that the CP symmetry is violated in the dark sector 
and both mixing terms are present in general. 

The U(1)$_{\rm d}$ gauge theory may be conformal in the presence of ``monopole" as well as ``electrons"~\cite{Argyres:1995jj, Argyres:1995xn}, 
which implies that the gauge field strength $F_{\mu \nu}$ has an scaling dimension larger than $2$ 
as is guaranteed by the unitarity bound~\cite{Mack:1975je}. 
As a result, the kinetic mixing terms are irrelevant operators 
and are suppressed at low energy~\cite{Kamada:2016ois}, if present. 
This naturally results in small $\epsilon'$ and $\epsilon$ in our model. 
Hereafter we represent $B_{\mu \nu}$ as the photon field strength and 
absorbs the Weinberg angle into $\epsilon'$ and $\epsilon$ for notational simplicity.

In this paper, we mainly consider the case with $\epsilon B_{\mu \nu} \tilde{F}^{\mu \nu}$ 
and without $\epsilon' B_{\mu \nu} F^{\mu \nu}$ for simplicity 
unless otherwise stated. 
In the dual basis, our model looks similar to the standard spontaneously broken U(1)$_{\rm d}$ gauge theory, where 
the U(1)$_{\rm d}$ symmetry is (spontaneously) broken by the condensation of the ``Higgs" field (i.e., the scalar ``monopole" in the original basis) 
and the kinetic mixing term looks the same as the usual one, $\epsilon B_{\mu \nu} F^{\mu \nu}$. 
Then we can quote constraints on the kinetic mixing parameter 
to compare our result with the present and future constraints. 
We will explain the case only with $\epsilon'$, which is qualitatively different but quantitatively similar to the case only with $\epsilon$.

Here we note that $\tilde{F}_{\mu\nu}$ does not satisfy the Bianchi identity, $\varepsilon^{\mu \nu \rho \sigma} \partial_{\nu} \tilde{F}_{\rho \sigma} = 0$, 
in theories consisting simultaneously of both a ``monopole" and an ``electron" (see, e.g., Ref.~\cite{Blagojevic:1985sh}). 
Then an operator mixing between 
$\tilde{F}_{\mu \nu}$ and ${\rm Tr} \left[ \pi \del_\mu \pi \del_\nu \pi \right]$ is allowed in those theories. 
Therefore, once we allow the nonzero kinetic mixing, $\epsilon B^{\mu \nu} \tilde{F}_{\mu \nu}$, 
we can have a term like 
\begin{equation}
 \mathcal{L} \supset c_\epsilon \frac{(4\pi)^2}{\Lambda^3} \epsilon 
 B^{\mu \nu} {\rm Tr} 
 \left[ \pi \del_\mu \pi \del_\nu \pi \right]. 
\label{kinetic mixing}
\end{equation}
where $c_\epsilon$ is an ${\cal O}(1)$ constant. 
This operator leads to a semi-annihilation process of $\pi \pi \to \pi \gamma$ only in the presence of a ``monopole" and ``electrons". 
If $\tilde{F}_{\mu \nu}$ satisfied the Bianchi identity, 
one could write $\tilde{F}_{\mu \nu} = \del_\mu \tilde{V}_\nu - \del_\nu \tilde{V}_\mu$ with $\tilde{V}_\mu$ being a (magnetic) gauge field of U(1)$_{\rm d}$. 
Then the kinetic mixing operator $B^{\mu \nu} \tilde{F}_{\mu\nu}$ 
could be written as $- 2 \del_\mu B^{\mu \nu} \tilde{V}_\nu = 0$ 
after the integration by parts for on-shell photon. 
However, 
$\tilde{F}_{\mu \nu}$ does not satisfy the Bianchi identity 
in the presence of a ``monopole" as well as ``electrons". 
There is no reason that we prohibit the operator of \eq{kinetic mixing} and the on-shell photon is produced by the annihilation process, $\pi \pi \to \pi \gamma$.

The operator of \eq{kinetic mixing} vanishes for $N_\pi < 3$ 
since it is antisymmetric in the flavor SU($N_F$), so that we assume $N_F \ge 2$ in our model. 
We note that the ``pions" transform as an adjoint representation of the flavor SU($N_F$). 
The two ``pions" in the initial state must be antisymmetric in the flavor SU($N_F$) to contact with the one ``pion" in the final state. 
On the other hand, the initial state of the semi-annihilation process 
must be symmetric in terms of the ``pion" exchange 
because ``pions" are bosons. 
These observations imply that the 
initial angular momentum must be antisymmetric 
and the semi-annihilation process is $p$-wave suppressed. 
We thus expect that its cross section can be estimated as 
\begin{equation}
 \left\langle \sigma v \right\rangle_{\pi \pi \to \pi \gamma}  \sim 
 c_\epsilon^2 \epsilon^2 \frac{(4\pi)^4 m_\pi^4}{4\pi \Lambda^6} \left( \frac{T}{m_\pi} \right), 
\end{equation}
from the dimensional analysis, where we absorb an ${\cal O}(1)$ uncertainty 
into $c_\epsilon$. 
This interaction is in thermal equilibrium at a temperature higher than $m_\pi$ 
for $c_\epsilon \epsilon \gtrsim 4 \times 10^{-12} \, c_{\Lambda}^{-3} (m_\pi / 100 \MeV)^{1/2}$. 
The temperature of the ``pions" is the same as that of the SM sector 
until the semi-annihilation process freezes out at $T / m_\pi \sim 1/20$.

\subsection{Relic abundance}

As the temperature becomes lower than the ``pion" mass, 
the number density of ``pions" is suppressed by the Boltzmann factor 
and eventually the $\pi \pi \to \pi \gamma$ semi-annihilation process freezes out. 
We note that the $\pi \pi \to \pi \gamma$ semi-annihilation process is similar to but is slightly different from 
the standard annihilation process in the weakly-interacting massive particle (WIMP) scenario. 
The important difference is that the ``pion" in the final state 
can be relativistic and may heat the dark sector~\cite{Kamada:2017tsq, Kamada:2017gfc, Kamada:2018hte}. 
From the Boltzmann equation of the ``pions", 
the evolution equations of the yield $Y_\pi$ ($\equiv n_\pi / s$) and the inverse temperature $x_\pi$ ($\equiv m_\pi / T_\pi$) are approximated as 
\beq
 &&\frac{d}{dx} Y_\pi \approx - \frac{\lambda}{x^2} Y_\pi^2, 
 \label{eq1}
 \\
 &&x \frac{d}{dx} \lmk \frac{x_\pi}{x} \rmk \approx \frac{x_\pi}{x} + \frac23 \bar{\lambda} Y_\pi 
 \lmk \frac{x_\pi}{x} \rmk^2, 
 \label{eq2}
\eeq
for $x$ ($\equiv m_{\pi} / T$) $> x_{\rm FO}$ ($\equiv m_{\pi} / T_{\rm FO}$), 
where $s = (2 \pi^2 / 45) g_* T^3$, 
$T$ is the temperature of the SM particles, 
and $T_{\rm FO}$ is the freeze-out temperature 
(see Ref.~\cite{Kamada:2017gfc} for the original equations without using approximations). 
The effective number of relativistic degrees of freedom, $g_*$, is taken to be about $10$. 
The dimensionless reaction rates are given by 
\beq
 \lambda = \frac{x s \la \sigma v_{\rm rel} \ra}{2 H}, 
 \quad
 \bar{\lambda} \approx - (\gamma -1) \lambda, 
\eeq
where $\gamma$ ($= 5/4$) is the Lorentz factor that DM achieves through semi-annihilation.

\begin{figure}[t]
\centering 
\includegraphics[width=.40\textwidth, bb=0 0 360 256]{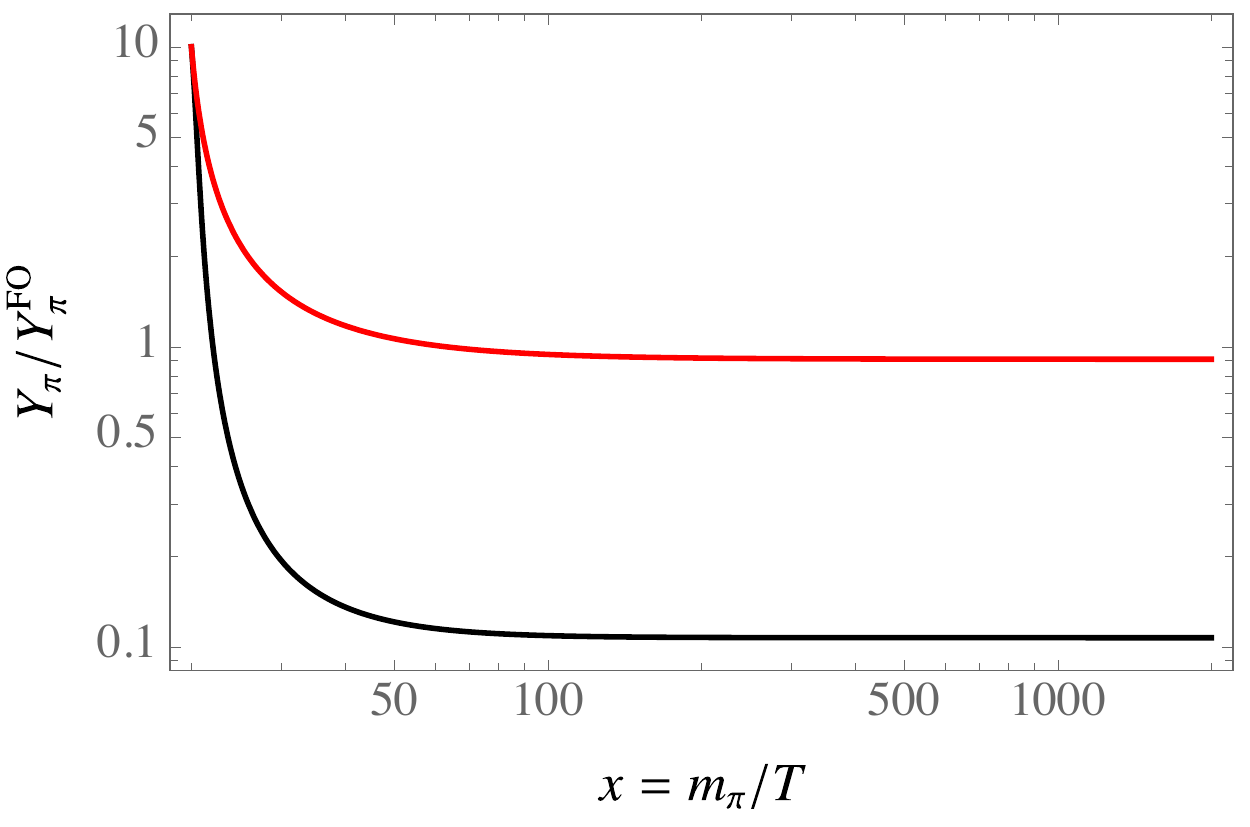} 
\\
\includegraphics[width=.40\textwidth, bb=0 0 360 256]{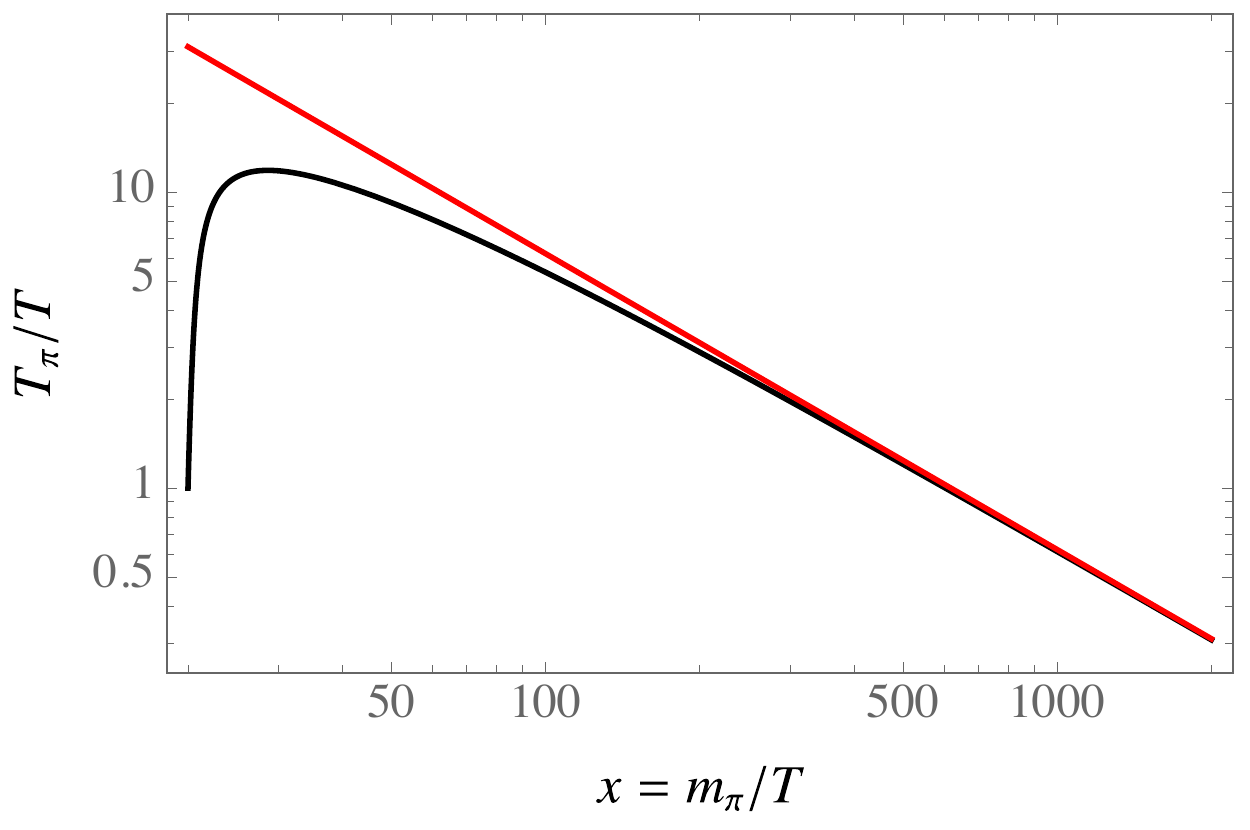} 
\caption{
Time evolutions of the yield $Y_\pi$ (black curve in the upper panel) and the temperature of ``pions" $T_\pi$ (black curve in the lower panel) 
for the case of $x_{\rm FO} = 20$. 
The red curve in the upper panel is the yield calculated in the case with $T_{\pi} = T$. 
The red curve in the lower panel is the asymptotic line of $30 T / T_{\rm FO}$. 
}
  \label{fig2}
\end{figure}

Assuming $x_{\rm FO} \sim 20$, 
we numerically solve Eqs.~(\ref{eq1}) and (\ref{eq2}). 
The time evolutions of the yield and the temperature of ``pions" are shown as black curves in Fig.~\ref{fig2}, 
where the yield is normalized by $Y_\pi^{\rm FO} \equiv 2 x_{\rm FO} / \lambda (x_{\rm FO})$. 
The red curve in the upper panel is the one without the self-heating 
while that in the lower panel is $x_\pi = 0.033 x^2 / x_{\rm FO}$ to which the numerical result asymptotically approaches. 
Thus we obtain 
the asymptotic value of the yield and the temperature of ``pions" as 
\beq
 Y_{\pi}^{\rm FO} \simeq c_Y \frac{2 x_{\rm FO}}{\lambda (x_{\rm FO})}, 
 \quad
 x_\pi \simeq c_x \frac{x^2}{x_{\rm FO}} , 
\eeq
for $x_\pi \gg x_{\rm FO}$, 
where $c_Y = {\cal O}(0.1)$ and $c_x = {\cal O}(0.1)$ are numerical constants.%
\footnote{
The initial condition is taken to be $Y_{\pi} = c_{\rm ini} x_{\rm FO}^2 / 
\lambda (x_{\rm FO})$ and $x_\pi = x$ at $x = x_{\rm FO}$ 
with $c_{\rm ini}$ being an ${\cal O}(1)$ constant. 
The numerical coefficients $c_Y$ and $c_x$ depend on $c_{\rm ini}$ only logarithmically 
while they linearly depend on $x_{\rm FO}^{-1}$. 
}
We note that there are ${\cal O}(1)$ uncertainties in these results, 
though they are accurate enough for our purpose. 
These results are different from the ones for the WIMP scenario 
by a factor of order $0.1$. 
This is because the relativistic ``pion" in the final state of the semi-annihilation process 
heats the dark sector, which results in the relative increase for the $p$-wave semi-annihilation rate. 
The energy density of the ``pions" at present 
is consistent with the observed value of the DM relic density 
when 
\begin{equation}
 \epsilon \sim 5 \times 10^{-7} c_Y^{1/2} c_\epsilon^{-1} c_\Lambda^{-3} 
 \left( \frac{m_\pi}{100 {\rm \ MeV}} \right). 
 \label{chipred}
\end{equation}
The kinetic mixing can be as large as, e.g., ${\cal O}(10^{-3})$ for $m_\pi = 100 \MeV$ 
if $c_\Lambda = c_\epsilon = 0.1$.

The second term in the right-hand side of \eq{eq2} 
becomes negligible after the freeze-out if the semi-annihilation process 
is $p$-wave suppressed and $\bar{\lambda} \propto 1/x_\pi$. 
Then the temperature of the ``pions" scales as $T_\pi \propto 1 /a^2$ just like the non-relativistic matter and the DM ``pions" is cold, 
where $a$ is the scale factor. 
This is in contrast to the case of a $s$-wave semi-annihilation process discussed in Ref.~\cite{Kamada:2017gfc}, 
where it is found that $T_\pi \propto 1/a$ 
because both the first and second terms in the right-hand side of \eq{eq2} are relevant and are balanced 
until the self-interaction freezes out. 
In the latter case, 
the temperature of DM is not that small and DM is warm, 
which is tightly constrained by measurements of the Lyman-$\alpha$ forest~\cite{Kamada:2018hte}. 
On the other hand, 
the temperature of DM decreases faster 
and DM is cold in our model.

We show the allowed region of the kinetic mixing parameter $\epsilon^2$ in Fig.~\ref{fig1}. 
We assume that $c_1, c_\Lambda, c_\epsilon \in (0.1 ,1)$ 
with a condition of $c_1 c_\Lambda^2 < (4 \pi)^{-1}$ (see footnote~\ref{footnote1}) 
for a conservative analysis 
while we take $c_Y =  0.1$ and $c_{m_V} = 1/4$ for simplicity. 
The shaded regions are parameters in which 
the DM relic abundance can be consistent with the observed DM abundance 
and the self-interaction cross section 
can be $\sigma_{\rm ela} / m_\pi \in (0.1, 1) \cm^2 / {\rm g}$. 
In the darkly shaded region, 
$\sigma_{\rm ela} / m_\pi$ can be as large as $1 \cm^2/ {\rm g}$ 
while in the lightly shaded region 
it is smaller than $1 \cm^2/ {\rm g}$ but can be larger than $0.1 \cm^2/ {\rm g}$. 
The upper-left corner of the shaded region is bounded by 
the condition that $c_\Lambda$ should not be smaller than about $0.1$ 
in \eq{chipred}. 
In the upper-right (lower-left) corner of the figure, 
the self-interaction cross section of ``pions" becomes too small (large) 
to be consistent with the observations of the small-scale structure. 
If $\epsilon^2$ is smaller than about $10^{-11}$ and $N_\pi \ge 5$, 
the $3 \to 2$ annihilation process becomes relevant during the freeze-out process 
as we will see shortly.

\subsection{Experimental constraints}

Since there is no $\pi$-$\pi$-$\gamma$ (or dark photon) interaction due to the flavor SU($N_F$), 
the ``pions" cannot be detected by the direct-detection experiments of DM. 
On the other hand, 
the dark photon can be produced via the kinetic mixing 
and can be discovered by some experiments employing missing momentum and/or energy techniques. 
In the figure, we plot the constraints on the kinetic mixing parameter 
by BaBar~\cite{Aubert:2008as, Izaguirre:2014bca, Lees:2017lec} and NA64~\cite{Banerjee:2016tad, NA64:2019imj} in the magenta and green lines, respectively. 
We can see that most of the parameter space is consistent with the present upper bound. 
The expected sensitivities of future experiments are shown by the dashed lines 
for Belle II (magenta)~\cite{Kou:2018nap, Inguglia:2019gub}, 
NA64 (green)~\cite{Beacham:2019nyx}, 
LDMX (blue), 
and Extended LDMX (red)~\cite{Akesson:2018vlm} experiments. 
We find that (Extended) LDMX experiment as well as Belle II experiment 
can cover a large parameter space.

\begin{figure}[t]
\centering 
\includegraphics[width=.40\textwidth, bb=0 0 360 256]{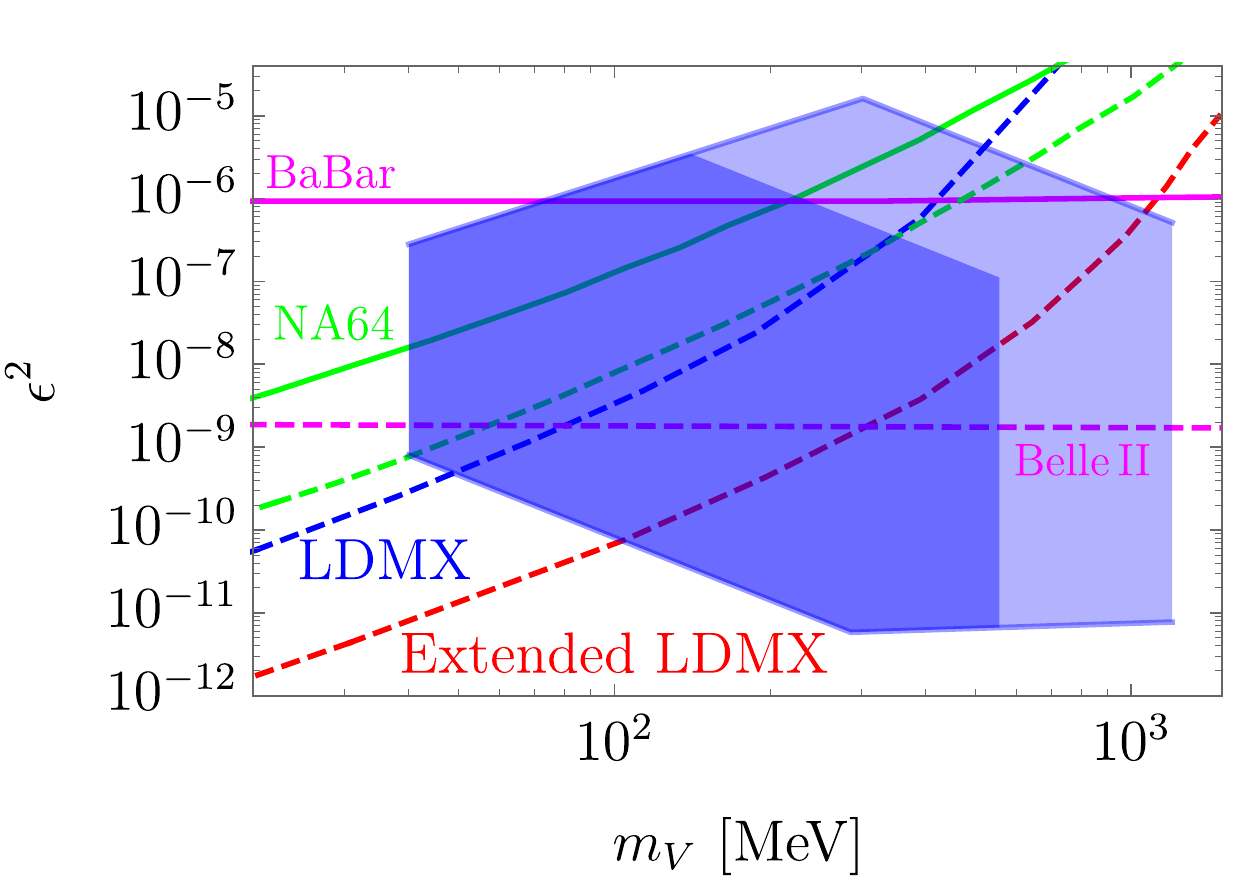} 
\caption{
Allowed region of the kinetic mixing parameter $\epsilon^2$. 
The shaded regions are parameters in which 
we obtain the correct DM relic abundance 
and $\sigma_{\rm ela} / m_\pi \in (0.1, 1) \cm^2 / {\rm g}$. 
The magenta and green lines are the upper bound by BaBar~\cite{Aubert:2008as, Izaguirre:2014bca} experiment and NA64~\cite{NA64:2019imj}, respectively. 
The dashed lines are the expected sensitivities of 
Belle II (magenta)~\cite{Kou:2018nap, Inguglia:2019gub}, 
NA64 (green)~\cite{Beacham:2019nyx}, 
LDMX (blue), 
and Extended LDMX (red)~\cite{Akesson:2018vlm} experiments. 
}
  \label{fig1}
\end{figure}

Note that the dark photon cannot decay into two ``pions" in our model. 
This implies that 
the dark photon cannot decay solely into the dark sector 
for the case of $m_V < 3 m_\pi$. 
On the other hand, 
the dark photon dominantly decays into the dark ``pions" 
for the case of $m_V > 3 m_\pi$. 
The LDMX experiment is designed to measure missing momentum in this kind of process. 
As we hope to indirectly detect the DM particle by LDMX-like experiments, 
we assume $m_V  = 4 m_\pi$  ($> 3 m_\pi$), i.e., $c_{m_V} = 1/4$, to plot the figure. 
We predict that $m_V$ is larger than about $30 \MeV$ 
because we require $m_V > 3 m_\pi$ and $m_\pi \gtrsim 10 \MeV$.

Here we comment on the case in which there is only the other kinetic mixing term $\epsilon' B^{\mu \nu} F_{\mu \nu}$ rather than $\epsilon B^{\mu \nu} \tilde{F}_{\mu\nu}$. 
In this case, 
\eq{kinetic mixing} should be replaced by a term like $c_{\epsilon'} (4 \pi)^2 / \Lambda^3 \epsilon \tilde{B}^{\mu \nu} \Tr [ \pi \del_\mu \pi \del_\nu \pi ]$ 
though our analysis of the semi-annihilation process does not change qualitatively. 
The standard model particles cannot emit on-shell dark photons 
while the dark-sector particles can be produced via the off-shell (dark) photons via the kinetic mixing. 
We expect that the cross section of such a process with missing particles 
is then given by the replacements of $m_V$ by $\Lambda$ and $\epsilon^2$ by $\epsilon'^2$ with an additional factor of $N_F \alpha_D / (2\pi) \ln (E / \Lambda)$ ($\sim {\cal O}(1)$) for $E \gtrsim \Lambda$, 
where $E$ ($= {\cal O}(1) \GeV$) is the energy of the scattering process~\cite{Izaguirre:2014bca}.%
\footnote{
One may think that the cross section is dominated by a low-energy contribution near the threshold of $3 m_\pi$~\cite{Izaguirre:2014bca}. In our case, however, it is negligible due to the $p$-wave suppression effect. 
}
We note that the additional factor is just an ${\cal O}(1)$ factor 
and the difference between $m_V$ and $\Lambda$ is also an ${\cal O}(1)$ factor. 
We may absorb these factors into $c_\epsilon$ and $c_{m_V}$, respectively. 
Then the result is similar to the one shown in Fig.~\ref{fig1} with $\epsilon^2 \to \epsilon'^2$. 
Even in the presence of both kinetic terms, 
the result does not change qualitatively because their effects are additive 
for the production process in the experimental setups as well as 
for the semi-annihilation process.

We also comment on the region near the lower bound on the ``pion" mass ($\sim 10 \MeV$). 
As the ``pions" are non-relativistic and are suppressed by the Boltzmann factor 
during the freeze-out process of neutrinos, 
the effect of ``pion" decoupling is almost negligible for observables such as the effective number of neutrinos. 
However, it is argued that its effect can be detected in the near future 
by the Simons Observatory~\cite{Ade:2018sbj} and CMB-S4~\cite{Abazajian:2016yjj, Abazajian:2019eic}
if the ``pion" mass is as small as about 10-15 MeV~\cite{Sabti:2019mhn}.

Finally, we note that the constraint from the indirect detection experiments of DM 
is not relevant in our model 
because the semi-annihilation process is $p$-wave suppressed 
and is not efficient in the galactic scale (see, e.g., Ref.~\cite{Boddy:2015efa}).

\subsection{$3 \to 2$ annihilation process}

The ``pions" may experience a $3 \to 2$ annihilation process 
via the following operator: 
\beq
c_{\rm WZW} \frac{(4\pi)^3}{N^{3/2} \Lambda^5} 
 \varepsilon^{\mu \nu \rho \sigma} \Tr \lkk \pi \del_\mu \pi \del_\nu \pi \del_\rho \pi \del_\sigma \pi \rkk. 
 \label{WZW}
\eeq
This term is allowed by any symmetry and is 
an analogy to the Wess-Zumino-Witten term in strong SU(N) gauge theories. 
It trivially vanishes for $N_\pi < 5$, namely $N_F < 3$. 
The cross section for the $3 \to 2$ annihilation process is calculated as~\cite{Hochberg:2014kqa}
\beq
 \la \sigma v^2 \ra_{3 \to 2} = 
 \frac{ (4 \pi)^6 c_{\rm WZW}^2 375 \sqrt{5} m_\pi^5}{2 \pi N_F \Lambda^{10}} \frac{T^2}{m_\pi^2}. 
\eeq
We should check that it is not efficient during the freeze-out of the $2 \to 2$ semi-annihilation process induced by \eq{kinetic mixing}. 
The condition is written as 
\begin{equation}
 \left\langle \sigma v^2 \right\rangle_{3 \to 2} 
 \left( n_\pi^{\rm eq} (T_{\rm FO}) \right)^2 
 \lesssim 
 \left\langle \sigma v \right\rangle_{\pi \pi \to \pi \gamma} 
 n_\pi^{\rm eq} (T_{\rm FO}) 
 \simeq 
 H(T_{\rm FO}). 
 \nonumber
 \\
\end{equation}
This condition is satisfied when 
\begin{equation}
 \epsilon \gtrsim 
 2 \times 10^{-6} c_\epsilon^{-1} \lmk \frac{c_{\rm WZW}}{0.1} \rmk^{3/5} 
 \lmk \frac{m_\pi}{100 \MeV} \rmk^{1/10}, 
 \label{lower bound on chi} 
\end{equation}
where we consider the case in which the relic abundance of ``pions" is consistent with the 
observed DM abundance. 
In Fig.~\ref{fig1}, 
the shaded region satisfies this condition 
with $c_{\rm WZW} = 0.1$ and $c_\epsilon =1$. 
However, we note that 
\eq{WZW} trivially vanishes 
and the constraint of \eq{lower bound on chi} is not applied 
for the case of $N_F = 2$ ($N_\pi = 3$), 
which is the minimal case for semi-annihilation to work in our model.

\subsection{Mixing between the SM Higgs and the ``monopole" 
\label{constraints}}

There must be a nonzero mixing between the ``monopole" $\phi$ and the SM Higgs field $H$ 
because the following interaction term is allowed by any symmetry: 
\beq
 V_{\rm mix} = \lambda \abs{\phi}^2 \abs{H}^2, 
\eeq
where $\lambda$ is a constant. 
After the Higgs and ``monopole" condensation, 
the mixing angle between the ``monopole" and the SM Higgs field is given by 
\beq
 \theta \simeq 0.023 \lambda c_{\rm mix} \lmk \frac{m_\pi}{1 \GeV} \rmk 
 \lmk \frac{m_V}{3 m_\pi} \rmk, 
\eeq
where we assume that the ``monopole"-condensation scale is related to $m_\pi$ by 
an ${\cal O}(1)$ factor $c_{\rm mix}$.

There is a strong collider constraint on the mixing parameter 
from the Higgs-decay channel into two ``monopoles"~\cite{Curtin:2013fra}. 
The ``monopoles" can decay into muons 
after they are produced from the Higgs decay~\cite{Ariga:2018uku}. 
In this case, 
the branching ratio of the Higgs decay into the ``monopoles" 
must be smaller than about $1\%$~\cite{Khachatryan:2015wka}, 
which requires that 
the quartic coupling $\lambda$ must be smaller than of order $10^{-3}$. 
Such a small coupling 
may be naturally realized in our model 
because our model may be conformal above the ``monopole" and ``electron" mass scale 
and the ``monopole" has a relatively large anomalous dimension~\cite{Argyres:1995jj, Argyres:1995xn}. 
The search for the Higgs decay into muons may also be an interesting direction to test our model 
in the near future.

\vspace{0.5cm} 

\section{Conclusions}
\label{conclusions}

We revisited our SIMP model with dark-sector ``electrons'' and a ``monopole" in U(1)$_{\rm d}$ gauge theory, 
motivated by the small-scale crisis in cosmology. 
We assumed ``monopole" condensation, which results in the formation of ``pions" in the low-energy sector. 
The relic abundance of the ``pions" is determined by the freeze-out process of semi-annihilation, $\pi \pi \to \pi \gamma$, that is induced from a kinetic mixing between the U(1)$_{\rm d}$ and U(1)$_Y$ gauge bosons. 
We note that on-shell photon can couple to the dark sector through the mixing with the U(1)$_{\rm d}$ gauge boson, since the U(1)$_{\rm d}$ field strength does not satisfy the Bianchi identity. 
The very kinetic mixing allows us to discover the U(1)$_{\rm d}$ gauge boson by LDMX-style missing momentum experiments in a large parameter space. 

We note that the model is quite economical: 
the U(1)$_{\rm d}$ gauge boson plays the roles of confinement 
and the mediator for the annihilation of ``pions". 
The number of flavour $N_F$ can be as small as two 
to introduce an operator for the semi-annihilation process. 
We assume SU($N_F$) flavour symmetry 
to ensure the stability of ``pions". 
One can promote 
this flavour symmetry to a gauge symmetry 
without changing our scenario qualitatively 
if the gauge coupling constant is small enough.

%
\vspace{0.2cm}
\begin{acknowledgments}
T.~T.~Y. deeply thanks the experimental groups at TDLI for the discussions on the search for the dark photon. Without the discussion, we could not have reached the conclusion in this paper.
We thank K. Yonekura for useful discussion. 
A.~K. was supported by Institute for Basic Science under the project code, IBS-R018-D1.
M.~Y. was supported by Leading Initiative for Excellent Young Researchers, MEXT, Japan. 
T.~T.~Y. was supported in part by the China Grant for Talent Scientific Start-Up Project and the JSPS Grant-in-Aid for Scientific Research No.~16H02176, No.~17H02878, and No.~19H05810 and by World Premier International Research Center Initiative (WPI Initiative), MEXT, Japan.
M.~Y. thanks the hospitality during his stay at DESY. 
T.~T.~Y. thanks Kavli IPMU for their hospitality during the corona virus outbreak. 
\end{acknowledgments}
%

\vspace{1cm}

\bibliography{references}

\end{document}